\documentclass[aps,prl,twocolumn,showpacs,groupedaddress,floatfix]{revtex4}

\usepackage{graphicx}
\usepackage{amsmath}
\usepackage{bbm}

\begin{document}
\title{XOR logic gate on electron spin qubits in quadruple coupled quantum dots}
\author{A. Kwa\'sniowski}
\affiliation{Faculty of Physics and Applied Computer Science, AGH
University of Science and Technology, Krak\'ow, Poland}
\author{J. Adamowski}
\affiliation{Faculty of Physics and Applied Computer Science, AGH
University of Science and Technology, Krak\'ow, Poland}

\date{\today}

\begin{abstract}
The spin-dependent localization of electrons in quadruple
quantum dots (QD's) has been studied by the configuration
interaction method.  We have investigated two nanodevices
that consist of laterally coupled quadruple QD's.
We have shown that -- in both the nanodevices with suitably chosen parameters --
the exclusive OR (XOR) logic gate can be realized by all-electrical control
with the readout of output via the spin-to-charge conversion.
We have determined the nanodevice parameters that are optimal for the
performance of the XOR logic gate.
\end{abstract}

\pacs{03.67.Lx,03.67.Mn,73.21.La}

\maketitle

Recent studies of quantum dot (QD) nanodevices towards a possible
application in quantum computation focus on a physical realization
of qubits and logic operations on them
\cite{Burk99,Hayashi03,Elz034,Sangu04,Petta05,Hans05,Hawr05,Meun06}.
In the QD's, the qubits can be encoded as two-level confined electron states
being either the orbital or spin states.
The orbital (charge) qubits are characterized by a short coherence
time \cite{Sauv02}, which is disadvantageous for a quantum
information processing. The spin qubits possess the coherence time
of the order of microseconds \cite{Hans03}, which makes them good
candidates for a storage and processing of quantum information. An
electrical manipulation of the electron qubits can been
realized in electrostatically gated QD's \cite{hand06}.
The results of the recent investigations of the laterally coupled multiple QD's
\cite{Hayashi03,Elz034,Petta05,Hans05,Hawr05,Meun06,Stopa06,Jiang08,Shin079}
are very promising.   In the double QD's, a coherent manipulation of electron spin qubits has been
demonstrated by Hayashi {\it et al.} \cite{Hayashi03} and Petta {\it et al.} \cite{Petta05}.
In the quadruple QD's, a controlled tunnelling and correlated charge oscillations
have been found by Shinkai {\it et al.} \cite{Shin079}.

Quantum computations can be performed using the elementary logic gates.
Barrenco {\it et al.} \cite{Bar95} showed that a set of one-qubit quantum gates and a two-qubit exclusive OR
(XOR) gate is universal, i.e., an arbitrary many-qubit gate can be decomposed
into these gates.  The possible realization of arbitrary one-qubit gate
in quantum-wire nanodevices has been proposed in Ref. \cite{BS089}  using the spin-orbit coupling and induced charge effects.
The simulations of the controlled NOT (CNOT) two-qubit gate have been performed
\cite{Moskal05} for electron charge qubits in coupled QD's.

In this Letter, we report on the results for the spin-dependent localization of electrons
in the nanodevices that consist of four laterally coupled gated QD's.
In these nanodevices, the potential confining the electrons can be tuned by changing the voltages
applied to the gates and/or source and drain electrodes.
We show that the quadruple QD nanodevices with suitably chosen parameters can perform the XOR logic gate
on electron spin qubits with the readout of output exploiting the spin-to-charge conversion.

\begin{figure}
\begin{center}
\includegraphics[width=0.2\textwidth,angle=270]{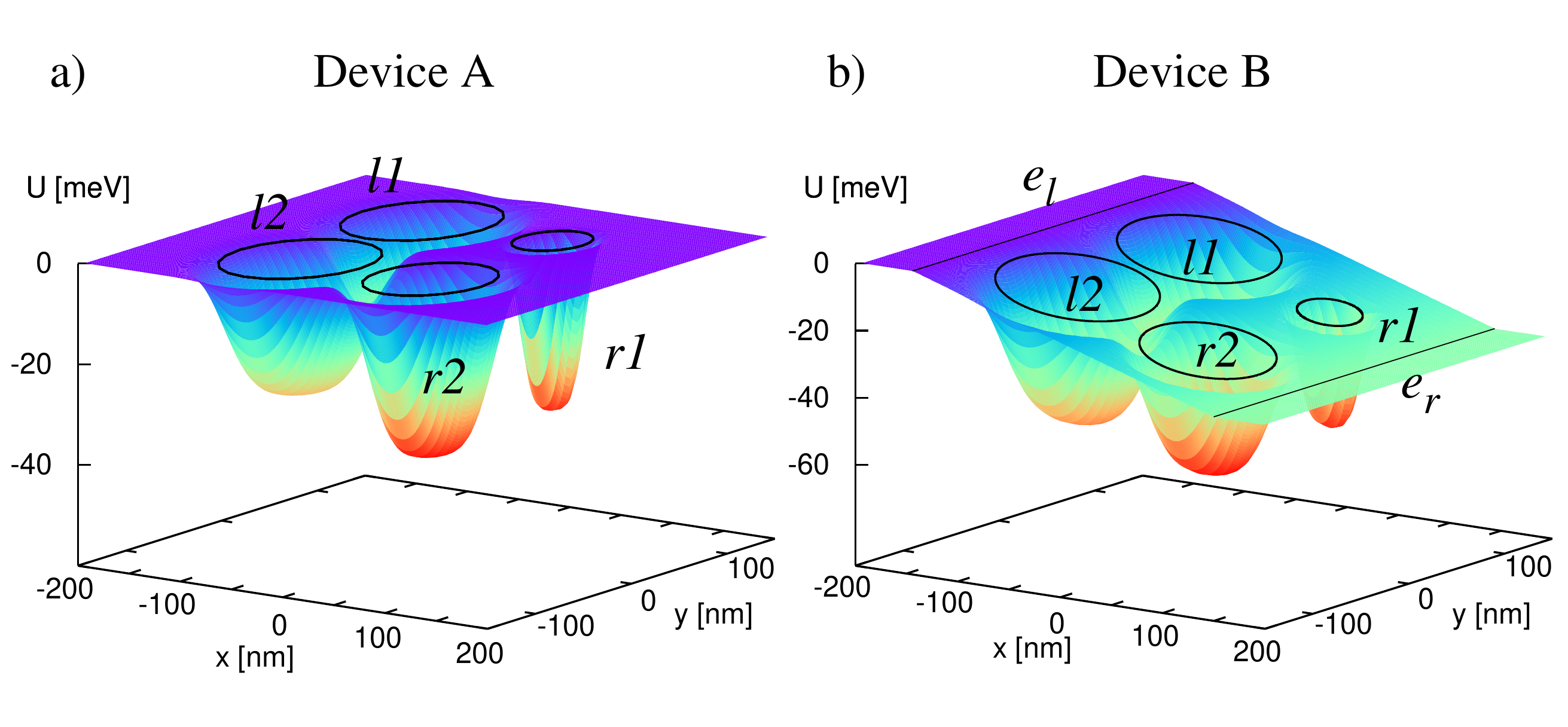}
\caption{(Color online) Potential energy $U$ of the electron in quadruple coupled QD's
as a function of $x$ and $y$ coordinates for device A (a) and B (b).
Ovals schematically show the sizes of the QD's labelled by $l1$, $l2$,
$r1$, and $r2$.  In device B, electric field $\mathbf{F} = (F,0,0)$
with $F$ = 1.01 kV/cm is applied.}
\label{fig1}
\end{center}
\end{figure}

In the quasi-two-dimensional quadruple QD \cite{Shin079}, the confinement potential energy of the electron
can be described by the following formula:
\begin{equation}
U_c(\mathbf{r}) = - \sum\limits_{\mu\nu} U^0_{\mu\nu}
\exp\{-[(\mathbf{r} - \mathbf{r}^0_{\mu\nu})^2/R_{\mu\nu}^2]^{p/2}\} \;,
\label{conf}
\end{equation}
where $\mathbf{r}=(x,y)$, the QD's are labelled by $\mu=l,r$ and $\nu=1,2$ (cf. Fig. 1), $U^0_{\mu\nu}$ is the depth
of the potential well ($U^0_{\mu\nu}>0$), $R_{\mu\nu}$ is the range of the confinement potential
that determines the size of QD$(\mu\nu)$,
$\mathbf{r}^0_{\mu\nu}$ is the position of the center of QD$(\mu\nu)$, and parameter $p \geq 2$ is responsible for the
softness of the confinement potential \cite{KA09}.
We consider two model nanodevices, denoted by A and B (Fig. 1).
In nanodevice A [Fig. 1(a)], the confinement potential energy
is given by Eq.~(\ref{conf}) with variable potential-well depths $U^0_{\mu\nu}$,
which can be tuned by changing the voltages applied to the nearby gates.
In nanodevice B [Fig. 1 (b)], the confinement potential
can be modified by switching on/off the bias voltage $V$ between electrodes $e_l$ and $e_r$.
Then, the electron gains the additional potential energy $\Delta U$.
Taking into account the finite range of electric field $\mathbf{F}=(F,0,0)$ \cite{KA09}
we put $\Delta U = -eV$ and $0$ at the right and left electrode, respectively,
and $\Delta U= -eF(x +L/2)$ in the region between the electrodes,
where $L$ is the distance between electrodes $e_l$ and $e_r$ and $F=V/L$.
The total potential energy $U=U_c+\Delta U$ of the electron as a function of $x$
and $y$ is displayed in Figs. 1 and 2.
In the initial state of the nanodevice, the electron potential energy of
the left QD's is set to be lower than that of the right QD's (cf. Fig. 2).

\begin{figure}
\begin{center}
\includegraphics[width=0.4\textwidth]{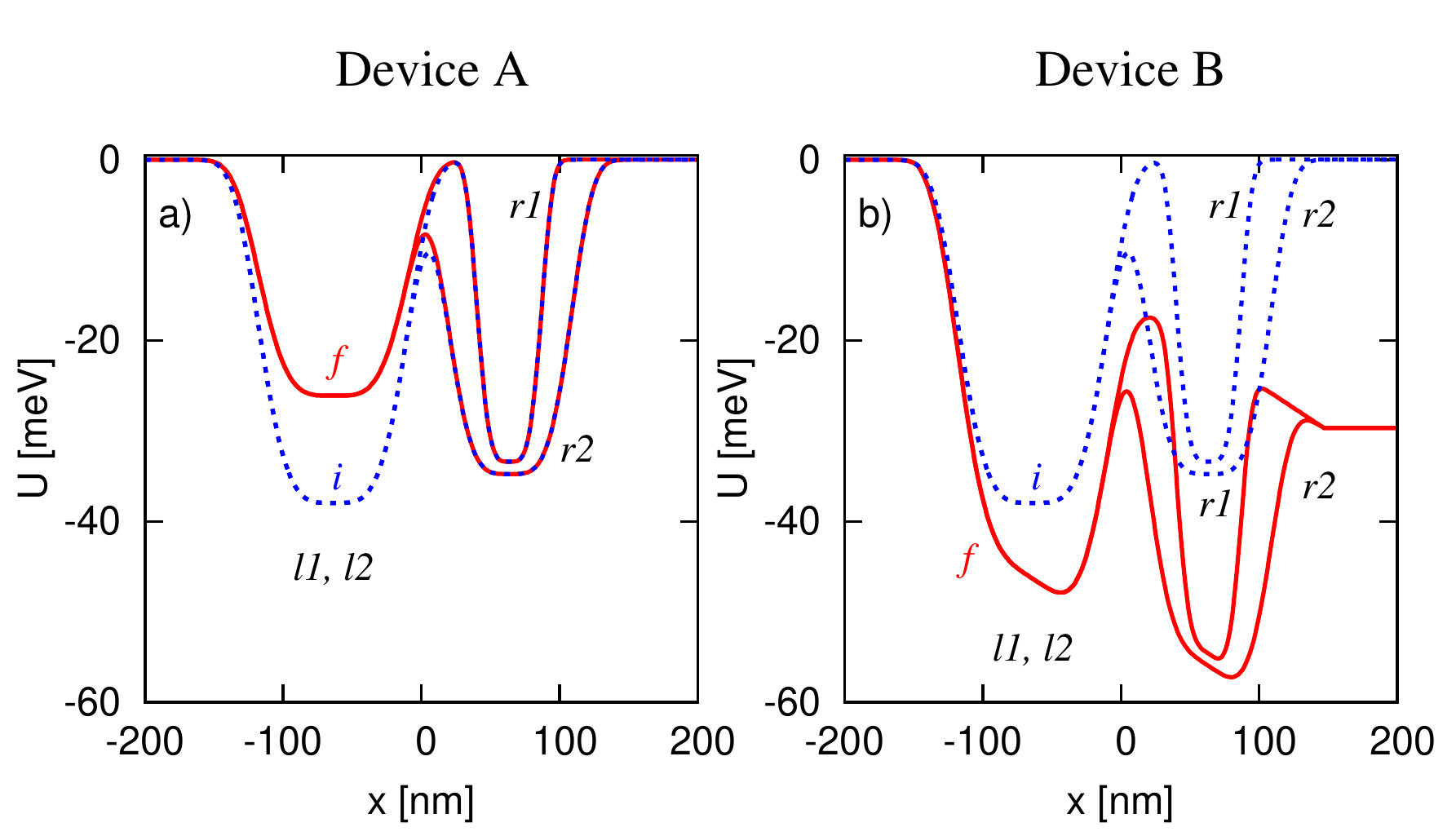}
\caption{(Color online)  Potential energy $U$ of the electron as a function of $x$ for $y$
fixed at the values corresponding to the straight lines joining the centers of QD$(l1)$-QD$(r1)$ and QD$(l2)$-QD$(r2)$
for device A (a) and B (b).  Dashed (blue) [solid (red)] curves show
the initial ($i$) [final ($f$)] potential energy profiles.}
\label{fig2}
\end{center}
\end{figure}

We have solved the Schr\"{o}dinger equation for one and two electrons in the quadruple QD
with the confinement potential energy $U$ (Fig. 1).
For this purpose we have extended the method applied previously to the double QD \cite{KA09}.
The two-electron problem has been solved by the configuration interaction method \cite{KA09}
on a space grid with the one-electron wave functions obtained
by the numerical variational method, described in Appendix of Ref. \cite{KA09}.
The computational method \cite{KA09} provides accurate solutions
to few-electron eigenproblems with an arbitrary confinement potential
and electric field of finite range.
The present calculations have been performed for the GaAs material parameters with donor rydberg
$\mathcal{R}=5.93$ meV and donor Bohr radius $a=9.79$ meV.
The initial potential energy is characterized by
$U^0_{l1}=U^0_{l2}=38$ meV, $R_{l1}=R_{l2}=58.7$ nm,
$U^0_{r1}= 33.4$ meV, $R_{r1}=25.5$ meV, $U^0_{r2}= 34.8$ meV, and $R_{r2}=49.0$ meV.
In the final state, $U^0_{l1}=U^0_{l2}=26.1$ meV for nanodevice A, while in nanodevice B,
the potential energy is changed by the electric field $F=1.01$ kV/cm.
The centers of QD's are located at the vertices of the square that are lying at distance 127.3 nm from each other.
The softness parameter is taken as $p=4$ [cf. Eq.~(\ref{conf})].
For each set of nanodevice parameters we have calculated the lowest singlet ($S$) and triplet ($T$)
energy levels for the initial and final potential energy profiles (cf. Figs. 1 and 2).
In the absence of magnetic field, the three triplet states ($T_0,T_{\pm}$) are degenerate,
therefore, we are dealing with the triply degenerate level $T$.
We have also determined the localization of electrons in the QD's for the lowest-energy $S$ and $T$ states
by calculating the one-electron probability density \cite{KA09}.

\begin{figure}[b]
\begin{center}
\includegraphics[width=0.25\textwidth,angle=270]{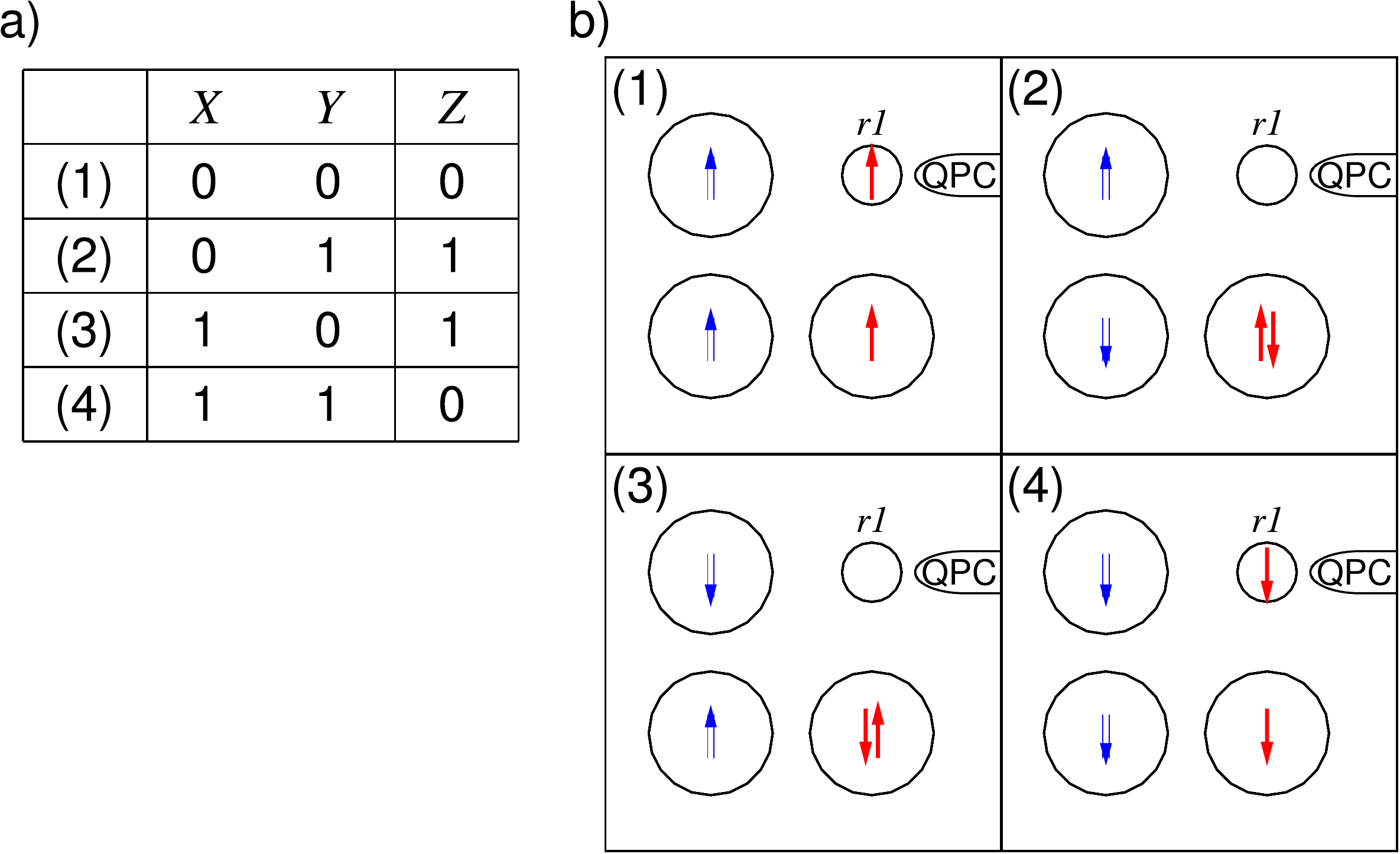}
\caption{(Color online) (a) Truth table for the XOR logic gate.
In columns $X$ and $Y$ ($Z$) the input (output) logical values are listed,
the numbers (1-4) in the left column identify the XOR gate operations.
(b) Schematic of the nanodevice with the localization
of electrons in QD's (circles) for operations (1-4).
The input logical value 0 (1) is encoded as the spin up (down) state
of the electron localized in the left QD.
The output logical values are defined as the charge states $q(r1)$
of QD$(r1)$ as follows: $q(r1) =0$ corresponds to
logical value 1 and $q(r1) = -e$ corresponds to logical value 0.
Charge $q(r1)$ is measured by the quantum point contact (QPC).
The spin states of the electrons are depicted by arrows: up
(down) arrow corresponds to the spin up (down).
The double (blue) [solid (red)] arrows schematically show the localization of electrons in the input
[output] states.}
\label{fig3}
\end{center}
\end{figure}

\begin{figure}[b]
\begin{center}
\includegraphics[width=0.35\textwidth,angle=270]{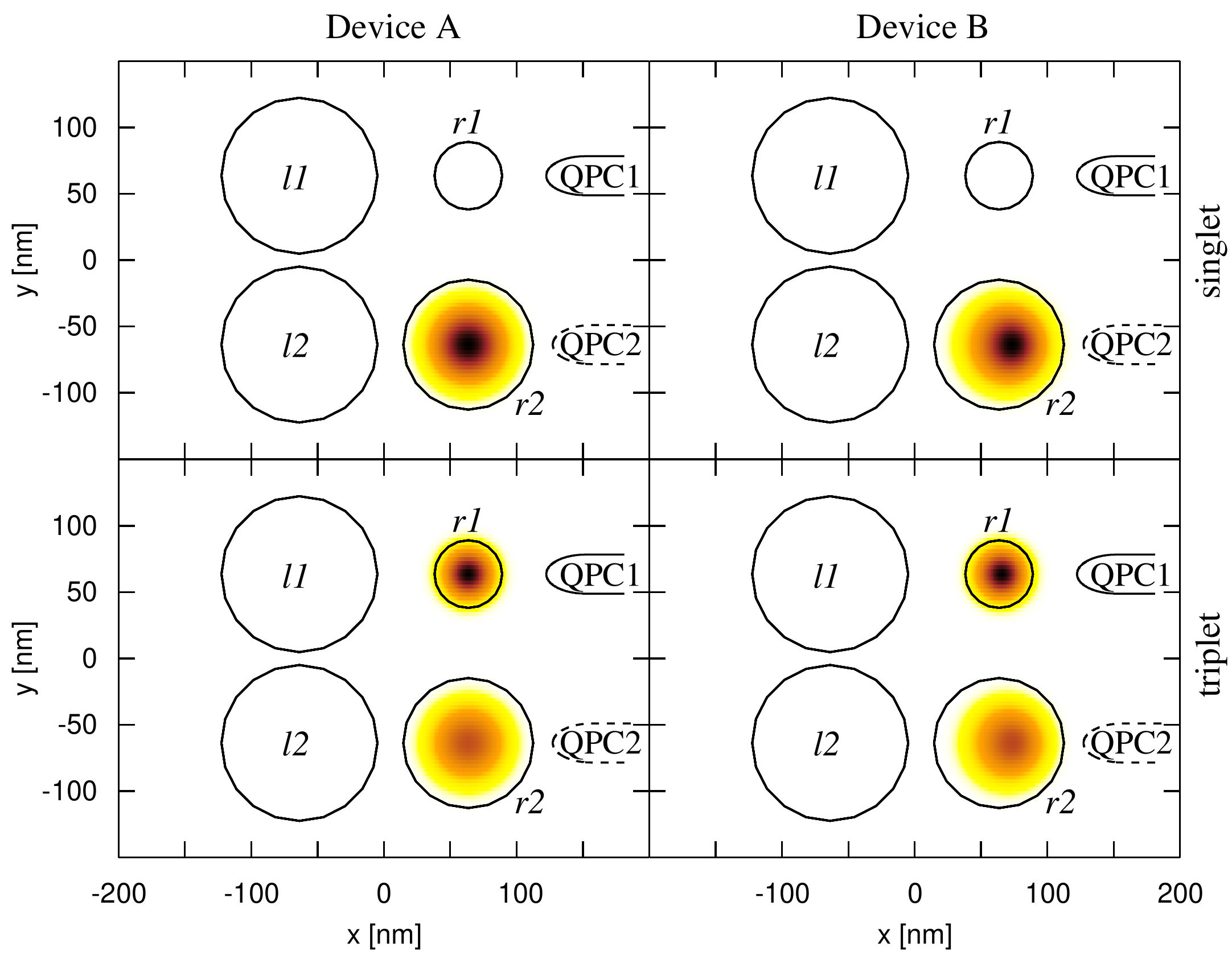}
\caption{(Color online)  Contours of one-electron probability density on the $x-y$ plane in the final state with
the electrons localized in the right QD's.  The nanodevice parameters are chosen so that
the XOR gate operations are realized in both the nanodevices: A (left panels) and B (right panels).
Circles correspond to the sizes of the QD's, QPC1 and QPC2 denote the quantum point contact in two alternative positions.
The two upper panels correspond to operations (2) and (3) (cf. Fig. 3) with the singlet-state
electrons localized in QD$(r2)$ and the two lower panels correspond to operations (1) and (4) with
the triplet-state electrons localized in QD$(r1)$ and QD$(r2)$.}
\label{fig4}
\end{center}
\end{figure}

The localization of two electrons obtained for the initial and final
potential profiles is schematically shown in Fig. 3 (b). Fig. 4 displays
contours of the one-electron probability density calculated for
the final potential profile (cf. Fig. 2). Fig. 4 shows that
the electron distribution in the single QD possesses the rotational
symmetry in nanodevice A and is slightly shifted towards the left
contact in nanodevice B, which results from the effect of the homogeneous electric
field.  The results of Fig. 4 can be analyzed
with the help of the truth table for the XOR logic gate [Fig.
3(a)].  The initial (input) state has been prepared so that
the left QD's are singly occupied by the electrons both in the singlet
and triplet states. This electron localization is ensured
by setting the potential wells of the left QD's to be deeper than
those of the right QD's (cf. Fig. 2). We ascribe the logical values 0 and 1 to the
initial spin states associated with the $z$ spin component
$+\hbar/2$ and $-\hbar/2$, respectively (Fig. 3). If we perform
the controlled evolution of the electron states by slowly changing
the potential energy profile (Fig. 2), the electrons become localized in
the right QD's (Fig. 4), since -- in the final state -- the potential wells of the right QD's are
considerably deeper than those of the left QD's.  The heights and thicknesses of energy
barriers separating the QD's have been chosen
sufficiently large to prevent the tunnelling of electrons between
QD$(l1)$ and QD$(l2)$ in the initial state.  During the changes of the potential energy profile, the energy
levels of the electrons confined in the left and right QD's become equal to each other,
which means that the resonant tunnelling conditions are satisfied, i.e.,
the electrons tunnel from the left to the right QD's  with
the probability $\sim$1. The occupancy  of the right QD's is determined by
the initial spin states of the electrons in the left QD's. If we consider the
QD$(r1)$, then -- for the nanodevice parameters given above --
this QD is occupied by either zero or one electron in the final state.
Simultaneously, the QD$(r2)$ is occupied by either
two electrons or one electron. This occupancy can be detected by
measuring charge $q(r1)$ of the QD$(r1)$ by a sensitive charge detector, e.g., the
quantum point contact schematically shown as QPC1 on Fig. 4.
Alternatively, we can use the quantum point contact QPC2, placed near the QD$(r2)$, which
measures charge $q(r2)$ of QD$(r2)$.  The results of the
present calculations allow us to predict the measurement outcomes.
For both the nanodevices A and B, we obtain either $q(r1)=0$ or
$q(r1)=-e$ and either $q(r2)=-2e$ or $q(r2)=-e$ for two different output
charge states (Fig. 4). This means that both the nanodevices A and
B operate as spin-charge converters. Moreover, these nanodevices
can act as the XOR logic gates (Fig. 3). When measuring the charge
of the single right QD, we can distinguish XOR logic
operations (1) and (4) from operations (2) and (3) [Fig. 3(a)]. If
we additionally know the initial electron spin states, we can
uniquely determine each of the four XOR logic gate operations
[Fig. 3(a)]. During operations (1-4) the input spin states
of the left QD's are transformed into the output charge states of the right QD's.

\begin{figure}[b]
\begin{center}
\includegraphics[width=0.4\textwidth]{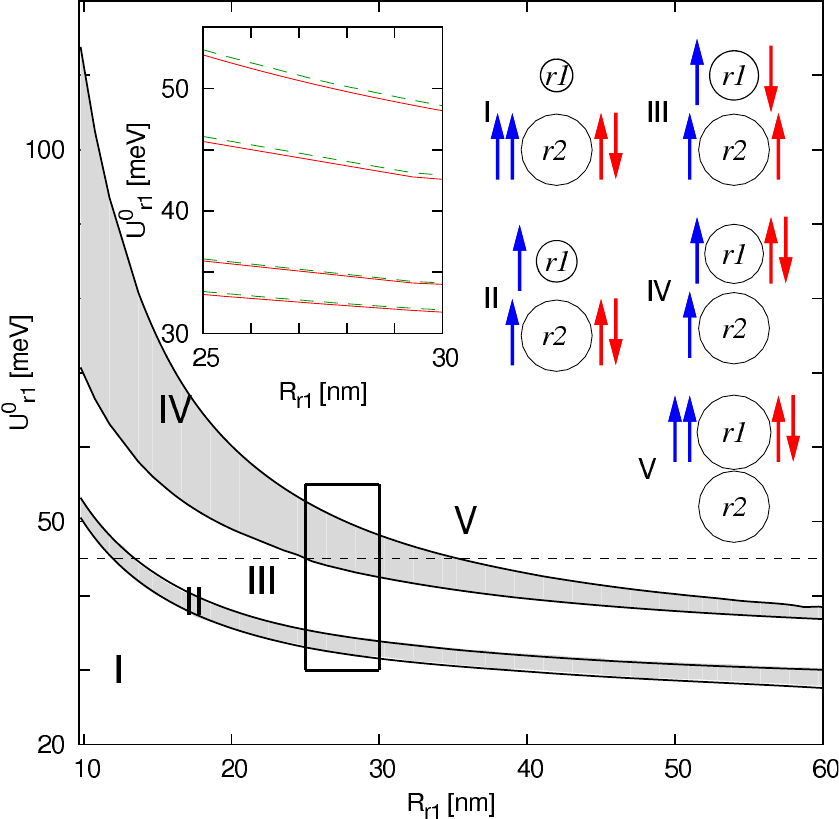}
\caption{(Color online) Boundaries of five regimes I-V (solid curves) of parameters $R_{r1}$ and $U^0_{r1}$
of QD$(r1)$.  Nanodevices A and B  with the parameters from regimes II and IV (grey)
can perform the XOR logic gate operations.
The right inset schematically shows the final-state localization of electrons
in the right QD's, with the parallel spins (blue arrows) and antiparallel spins (red arrows).
The horizontal dashed line allows us to trace how the electron localization is varied when passing
between regimes I-V by changing the size of QD$(r1)$.
The left inset is the zoom of the rectangular region shown on the main panel: solid (red) [dashed (green)] curves display
the boundaries of regimes I-V obtained for device A [B].}
\label{fig5}
\end{center}
\end{figure}

The nanodevices A and B can perform the XOR logic gate only for the suitably chosen nanodevice parameters.
In order to determine the corresponding parameter regimes,
we have performed many computational runs for different sets of parameters of QD$(r1)$
keeping fixed the parameters of other QD's.
The localization of electrons in the right QD's obtained for the final potential profile
is schematically depicted in the right inset of Fig. 5.
Fig. 5 shows that -- based on the criterion of electron localization in the final state --
we can distinguish five regimes, labelled by I through V, on the $R_{r1}-U^0_{r1}$ plane.
In regime I, both the electrons are localized in the QD$(r2)$ in either spin state.
In regime II, we obtain the spin-dependent localization of the electrons in QD$(r1)$ and QD$(r2)$.
The nanodevices A and B with the parameters from regime II can operate as the XOR logic gates (cf. Fig. 4).
In regime III, the electrons in each final spin state are localized in different QD's.
In regime IV, we recover the spin-dependent localization of the electrons in the right QD's.
Therefore, the nanodevices A and B with the parameters taking on the values from regime IV
can perform the XOR logic gate operations.
In comparison to regime II, the role of QD$(r1)$ and QD$(r2)$ is interchanged, which means that -- in regime IV --
the QPC2 will provide the same measurement outcomes as the QPC1 in regime II and {\it vice versa}.
In regime V, both the electrons occupy the QD$(r1)$ in each final spin state.

The left upper inset in Fig. 5 displays the zoom of the rectangular region on the main panel
with the boundaries of regimes I-V obtained for nanodevices A and B.  We observe that
the boundaries of these parameter regimes change only slightly (these changes are of the order of
the thickness of curves on the main panel).
The smallness of these changes means that both the nanodevices A and B are essentially equivalent
in the proposed realization of the XOR logic gate.

If the initial spin state is prepared as schematically shown in Fig. 3(b), i.e.,
the left QD's are singly occupied by the electrons with either parallel or antiparallel spins,
then the controlled quantum-state evolution performed by changing the electrostatic potential profile
leads to the uniquely determined final quantum states of electrons.  This occurs in all parameter regimes I-V (Fig. 5).
In the final state, the electrons are localized in either one or two right QD's in
the well-defined spin states (cf. the right inset of Fig. 5).  However, only in regimes II and IV, the
localization of electrons in QD$(r1)$ and QD$(r2)$ depends on their spins.
For the nanodevices A and B, characterized by the parameters
from regimes II and IV, the measurements of the electric charge of either QD$(r1)$ or QD$(r2)$
allow us to distinguish the singlet and triplet final spin states (Fig. 4).
Therefore, the nanodevices A and B can be used to the electrical readout of spin
via the spin-to-charge conversion.

The present mechanism of the XOR logic gate is based on the results obtained for the stationary quantum states. In order
to perform the controlled evolution between these states we have to apply the
external electric field that appropriately modifies the electron potential
energy.  The electric field has to be switched on/off sufficiently
slowly in order to cause the adiabatic transition from the initial to final state.
The non-adiabatic transitions can lead to several spin flips during the
quantum-state evolution, which makes the operation time longer
\cite{Moskal07}.  Nevertheless, the present mechanism leads to the realization of the XOR gate, if
the spins of electrons are the same in the initial and final states [cf. Fig. 3(b)].
The electrons only change their localization as a result of tunnelling from the left to right  QD's.
Therefore, the XOR gate operation time may be short enough in order to
perform the sufficiently large number of operations during the
spin coherence time \cite{Hans03}.

We note that the two-qubit XOR logic gate is commonly defined as
follows: $|X\rangle|Y\rangle \longrightarrow |X\rangle |Z=X\oplus
Y\rangle$, where $\oplus$ is the addition modulo two.  This
two-qubit XOR gate is equivalent to the CNOT gate \cite{DiV98}.  However,
according the present mechanism, the nanodevices A and B perform the
following operation: $|X\rangle_{l1}|Y\rangle_{l2}
\longrightarrow |Z=X\oplus Y\rangle_{r1}$, where $X, Y$, and $Z$
are defined in Fig. 3(a) and the output is detected by QPC1 as the charge of QD$(r1)$ (cf.
Fig. 4).  If we alternatively apply the QPC2 to measure the charge of QD$(r2)$,
then the final charge state $|Z=X\oplus Y\rangle_{r2}$
is also uniquely determined by the input spin states of the left
QD's.   Therefore, the logic gate studied in this Letter
possesses all the properties of the XOR gate defined in Fig. 3(a). The logic gate operations
realized in nanodevices A and B result from the
quantum transitions between the well-defined quantum states of
the electrons.  During these operations, the input spin
qubits are transformed into the output charge qubits by changing the
external voltages, i.e., by the all-electrical control.
The nanodevices A and B can realize the entire cycle
of the XOR logic gate, i.e., the input state preparation, electrically driven evolution of quantum states,
and readout of the output via the spin-to-charge conversion.
The XOR logic gate, proposed in this Letter,
can be realized in the quadruple QD's recently studied by Shinkai {\it et al.} \cite{Shin079}.

In summary, we have proposed the all-electrical implementation of the XOR logic gate in two
nanodevices based on the laterally coupled quadruple QD's.

\acknowledgments

This paper has been partly supported by the Polish Scientific Network LFPPI
''Laboratory of Physical Fundamentals of Information Processing''.

\end{document}